\documentclass{appolb}
\usepackage{graphicx}
\begin{document}
\title{From chiral NN(N) interactions to giant and pygmy resonances via extended RPA%
\thanks{Presented at the Zakopane Conference on Nuclear Physics “Extremes of the Nuclear
Landscape”, Zakopane, Poland, August 28 $–$ September 4, 2016}%
}
\author{
       Panagiota Papakonstantinou$^1$, Richard Trippel$^2$,  Robert Roth$^2$  
\address{$^1$Rare Isotope Science Project, Institute for Basic Sceince,
Daejeon 34047, S.Korea\\
           $^2$Institut f\"ur Kernphysik, T.U. Darmstadt, 64283 Darmstadt, Germany 
       }
           }
\maketitle
\begin{abstract}
The properties of giant and pygmy resonances are calculated starting from chiral two-and three-nucleon interactions. The aim is to assess the predictive power of modern Hamiltonians and especially the role of the three-nucleon force. Methods based on the random-phase approximation (RPA) provide an optimal description of the modes of interest with minimal computational requirements. Here we discuss the giant resonances (GRs) of $^{40,48}$Ca isotopes and their low-energy dipole response. A comparison with previous results obtained with a transfromed Argonne V18 two-nucleon potential points to certain improvements.  
\end{abstract}
  
\section{Introduction}

A starting point for nuclear structure theory ideally involves realistic two-plus-three nucleon (NN+NNN) potentials and, most consistently, nuclear Hamiltonians derived from quantum chromodynamics. Starting from these interactions, unitary transformations can be employed, e.g. the Similarity Renormalization Group (SRG), to pre-diagonalize the Hamiltonian and to improve the convergence behavior of many-body methods. This approach has been applied successfully to light and medium-mass nuclei using interactions from chiral effective field theory ($\chi$EFT) in the framework of the No-Core Shell Model and in Coupled-Cluster Theory and related methods.

In order to reach computationally heavy nuclei as well as higher-lying collective excitations, we have been exploring the performance of pre-diagonalized interactions within the Random Phase Approximation (RPA) and extensions thereof, in particular the Second RPA (SRPA). A two-body Hamiltonian based on the Argonne V18 potential was used before in large-scale SRPA calculations~\cite{PaR2009}  with promising results for giant resonances (GRs), notwithstanding the insufficient treatment of three-body effects. 
Since then, NN+NNN $\chi$EFT interactions have become available~\cite{MaE2011,Rot2012}. They can be utilized in a two-body formalism, by performing a normal ordering and neglecting the three-nucleon residual interaction - a truncation whose validity has been demonstrated~\cite{Rot2012}. Thanks to the above advances, we are now in a position to study collective phenomena 
with realistic potentials and with reasonable computational effort. 

As a linear-response theory, RPA would be the obvious many-body method of choice. 
There are two main reasons to go beyond first-order RPA, SRPA: The traditional phenomenologist$’$s goal is to describe the resonances' fragmentation because of collisional damping. The many-body theorist$’$s goal applicable here is convergence with respect to the model space, when the functional or interaction is not fitted to the mean field and RPA level. 

This contribution 
focuses on recent results within RPA and SRPA~\cite{TPR2016XX} and the relevance of utilizing a realistic three-nucleon interaction. 

\section{Giant resonances} 

In the past we employed the SRPA with the Argonne V18 potential transformed via the unitary correlation operator method (AV18+UCOM) and looked at the giant monopole resonance (GMR), giant dipole resonance (GDR), and giant quadrupole resonance (GQR)~\cite{PaR2009}.  Within in a 15-shell model space, a very good and almost-converged description of the GDR and GQR was obtained, including some very interesting applications in the observed fragmentation of the GQR~\cite{Usm2011,Usm2016}, but the energy of the GMR was underestimated. Overal, the energetic discrepancies were between, approximately, -10 (GMR) to 0 (GQR) MeV. The calculated charge radii also were too small. We attributed the discrepancies to missing NNN effects. 

One is now in a position to use NN+NNN interactions determined in a systematic way. The interaction used at present is a chiral NN (at N$^3$LO) and NNN (at N$^2$LO) interaction with a cutoff at 400~MeV, evolved within the SRG ($\chi$EFT+SRG). The NNN interaction is rewritten in a normal ordered form. The one and two body operators are kept, whereas the NNN residual interaction is neglected. Then the convenient two-nucleon formalism can be used. 

Fig.~1(a),(b) shows results for the GRs of the $^{40,48}$Ca isotopes obtained within a 13-shell model space.
Previous results with the AV18+UCOM potential in the same space but within SRPA0 are shown for comparison, as well as experimental data. SRPA0 stands for the diagonal approximation~\cite{PaR2009}, whereby the couplings amongst the $2p2h$ configurations are neglected. It has been found very good whenever tested against full SRPA. The energetic discrepancies, with respect to data, observed with the two potentials are of a different quality: When using the AV18+UCOM the energies are underestimated, while with the use of the $\chi$EFT+SRG they are overestimated. The latter results could be ameliorated still if we extend the harmonic-oscillator basis. The new results on GRs therefore constitute an improvement with respect to AV18+UCOM. However, the radii are still too small. In particular, the obtained values for the root-mean-square charge radii of $^{16}$O, $^{40}$Ca and $^{48}$Ca are 2.41, 2.98 and 2.6~fm, to be compared with the measured values 2.70, 3.48 and 3.48~fm, respectively. Next we may consider other versions of chiral interactions, for example the SAT family~\cite{Eks2015}, or the new two-body Daejeon16 interaction~\cite{Shi2016}, which promise improved radii.

\section{On the low-energy dipole spectrum} 

Another interesting benchmark, especially because it is qualitative, is the low-energy isovector (IV) dipole response of Ca isotopes and the nature of the low-energy isoscalar state (IS-LED)~\cite{PHP2012,Der2014}. The AV18+UCOM potential yields extremely strong low-energy transitions for $^{40,48}$Ca and predicts a neutron-skin oscillation for $^{48}$Ca. The result qualitatatively contradicts observations~\cite{Der2014}. Simple phenomenological corrections~\cite{GPR2014} could not improve this stiff result. Numerical results for the IS-LED in Ca with the new $\chi$EFT+SRG NN+NNN interaction are shown in Fig.~1(c). The new results constitute an improvement of an order of magnitude. Furthermore the IS-LED is not predicted to be a veritable neutron-skin oscillation in $^{48}$Ca. Whether the properties of a realistic NNN force are responsible for this outcome will have to be investigated using different NN(+NNN) potentials and different test nuclei.

\section{Prospects} 
The correct desription of giant and pygmy resonances is a potential benchmark for chiral and other modern nuclear interactions. 
Results with a chiral NN+NNN potential are promising. Next we shall examine the SAT family of chiral interactions and the two-nucleon Daejeon16 interaction. In the spirit of {\em ab initio} nuclear structure, we aim for a theoretical description of nuclear linear response based on chiral and in general microscopic interactions, for more-unbiased results and predictive power. 

\vspace{2mm}
\noindent 
{\bf Acknowledgements} PP's work is supported by the Rare Isotope Science Project of the Institute for Basic Science funded by Ministry of Science, ICT and Future Planning and the 
National Research Foundation (NRF) of Korea (2013M7A1A1075764).
RT's and RR's work is supported by DFG through the SFB 1245 and BMBF through contract 05P15RDFN1.

\begin{figure}[htb]
\mbox{$~$}\hspace{10mm}(a)   \hspace{30mm} (b) \hspace{30mm} (c)  \\[-2mm] 
\includegraphics[width=8cm]{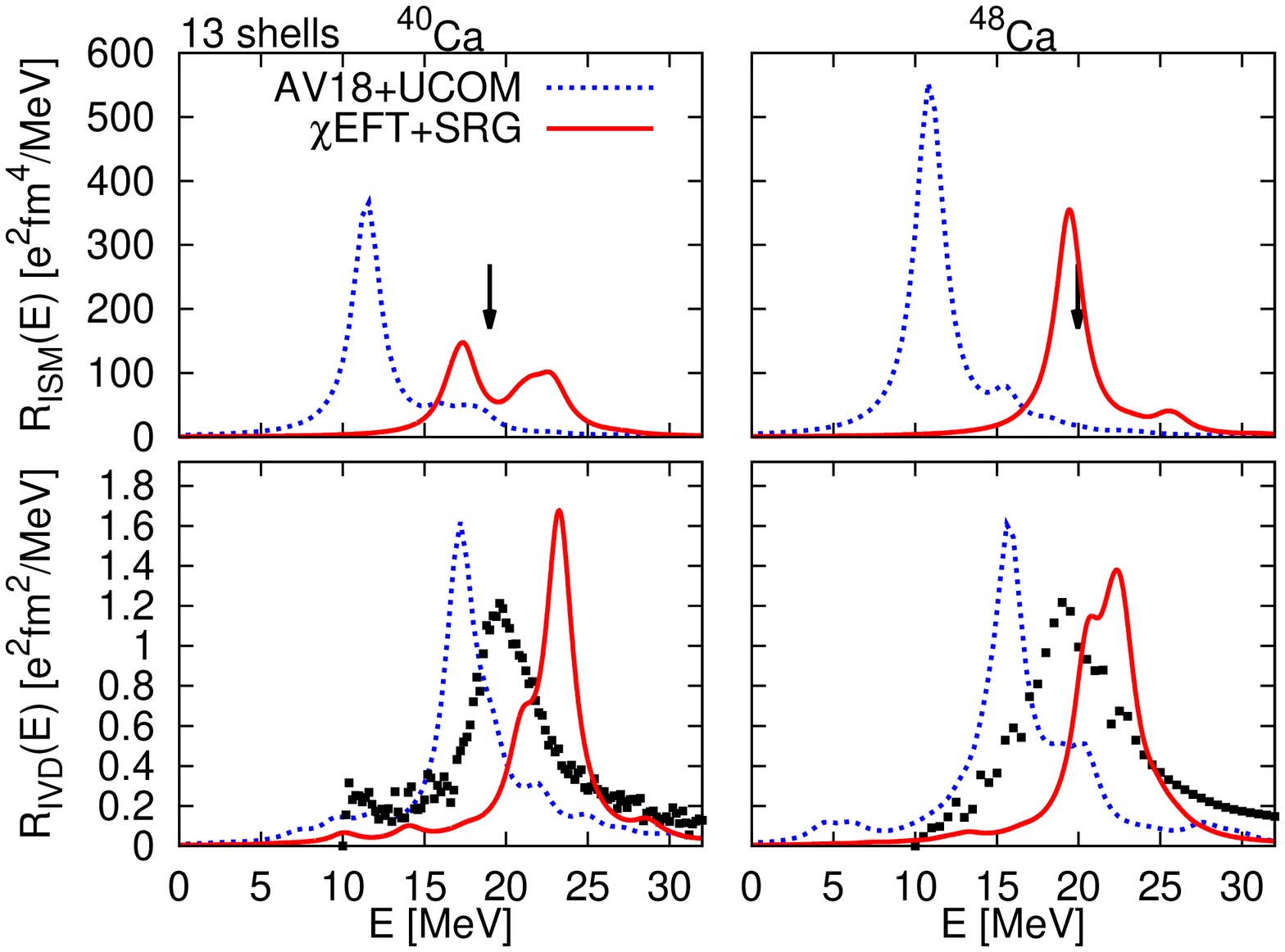}
\includegraphics[width=4cm,height=5.8cm]{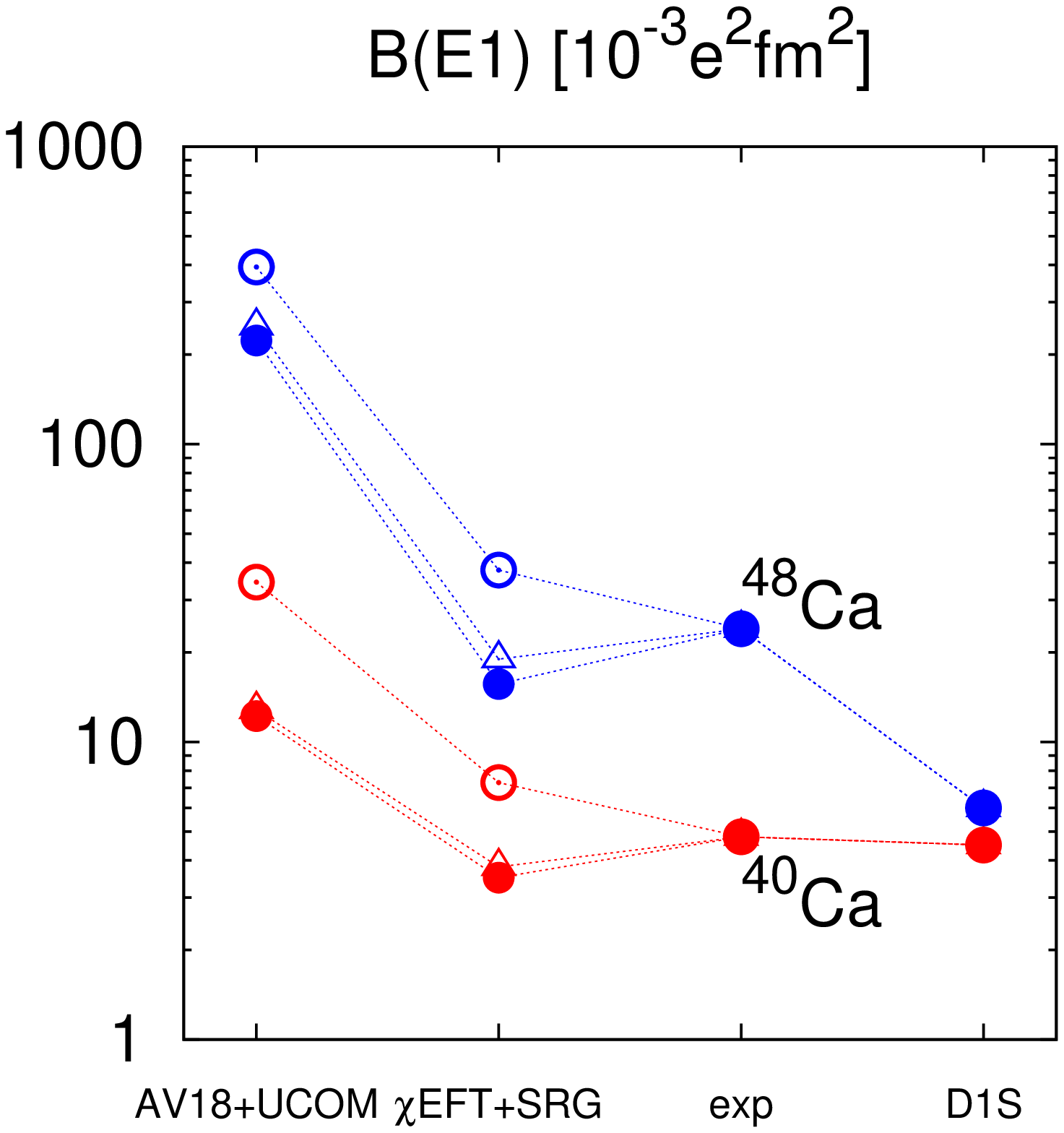}
\caption{(a),(b): The isoscalar monopole (ISM) and the isovector dipole (IVD)
strength distributions of $^{40,48}$Ca calculated using the AV18+UCOM or the $\chi$EFT+SRG interaction. Measured centroid GR energies and photoabsorption data are displayed. (c) $B(E1)\!\!\uparrow$ strength of the IS-LED. AV18+UCOM and $\chi$EFT+SRG: full points show the SRPA results, open symbols show the SRPA0 (triangles) and RPA results. Gogny D1S: only RPA. The experimental point for $^{48}$Ca includes the IV state in the immediate proximity of the IS-LED~\cite{Der2014}. 
}
\label{Fig:F2H}
\end{figure}


\begin{thebibliography}{10}

\bibitem{PaR2009}
P.~Papakonstantinou, R.~Roth,
\newblock {\em Phys. Lett.}
\newblock {\bf  B671}, 356 (2009).

\bibitem{MaE2011}
R.~Machleidt, D.R. Entem,
\newblock {\em Physics Reports}
\newblock {\bf  503}, 1  (2011).

\bibitem{Rot2012}
R.~Roth, S.~Binder, K.~Vobig, A.~Calci, J.~Langhammer, P.~Navr\'atil,
\newblock {\em Phys. Rev. Lett.}
\newblock {\bf  109}, 052501 (2012).

\bibitem{TPR2016XX}

\newblock R.~Trippel, P.~Papakonstantinou, R.~Roth (in preparation).

\bibitem{Usm2011}
I.~Usman, et~al.,
\newblock {\em Phys. Lett. B}
\newblock {\bf  698}, 191 (2011).

\bibitem{Usm2016}
I.~T. Usman, et~al.,
\newblock {\em Phys. Rev. C}
\newblock {\bf  94}, 024308 (2016).

\bibitem{Eks2015}
A.~Ekstr\"om, et~al.,
\newblock {\em ibid.}
\newblock {\bf  91}, 051301 (2015).

\bibitem{Shi2016}
A.M. Shirokov, I.J. Shin, Y.~Kim, M.~Sosonkina, P.~Maris, J.P. Vary,
\newblock {\em Phys. Lett. B}
\newblock {\bf  761}, 87  (2016).

\bibitem{PHP2012}
P.~Papakonstantinou, H.~Hergert, V.Yu Ponomarev, R.~Roth,
\newblock {\em ibid.}
\newblock {\bf  709}, 270 (2012).

\bibitem{Der2014}
V.~Derya, et~al.,
\newblock {\em ibid.}
\newblock {\bf  730}, 288 (2014).

\bibitem{GPR2014}
A.~G\"unther, P.~Papakonstantinou, R.~Roth,
\newblock {\em Journal of Physics G: Nuclear and Particle Physics}
\newblock {\bf  41}, 115107 (2014).

\end{thebibliography}
\end{document}